\documentclass[aps,prc,twocolumn,amsmath,amssymb,floatfix,superscriptaddress]
{revtex4-1}
\usepackage{mathptmx}
\usepackage[T1]{fontenc}
\usepackage{CJK}
\usepackage{graphicx}
\usepackage{bm}
\usepackage{dcolumn}
\usepackage{xcolor}
\usepackage{epstopdf}
\usepackage{multirow} 
\usepackage{booktabs} 
\usepackage[normalem]{ulem}
\usepackage{xcolor}
\DeclareRobustCommand{\erase}{\bgroup\markoverwith{\textcolor{red}{\rule[.5ex]{2pt}{0.4pt}}}\ULon}

\usepackage[
breaklinks,pdfstartview=FitH,CJKbookmarks=true,
bookmarksnumbered=true,bookmarksopen=true,pdfborder={0 0 1},
colorlinks=true,linkcolor=blue,urlcolor=blue,anchorcolor=blue,citecolor=blue
            ]{hyperref}

\def\be{\begin{equation}} \def\ee{\end{equation}}
\def\bal#1\eal{\begin{align}#1\end{align}}
\def\bse#1\ese{\begin{subequations}#1\end{subequations}}

\def\eps{\varepsilon}
\def\la{\Lambda}
\def\rv{\bm{r}}
\def\Jv{\bm{J}}
\def\Wv{\bm{W}}
\def\Nv{\bm{\nabla}}
\def\ron{\rho_N}

\def\rol{\rho_\la}

\def\fmq{\,\text{fm}^{-3}}

\def\mfm5{\;\text{MeV}\,\text{fm}^5}

\begin{document}

\begin{CJK*}{UTF8}{gbsn}
\title{Charge symmetry breaking effect in mirror $\Lambda$ hypernuclei with Skyrme-Hartree-Fock model}

\author{Shi-jing~Zha (查士婧)}
\affiliation{
	Department of Physics, East China Normal University,
	Shanghai 200241, China}
\author{Suo Qiu (裘索)}%
\affiliation{
	Department of Physics, East China Normal University,
	Shanghai 200241, China}

\author{H.~Sagawa}
\affiliation{
Center for Mathematics and Physics, University of Aizu, Aizu-Wakamatsu, Fukushima 965-8560, Japan}
\affiliation{
RIKEN, Nishina center for Accelerator-Based Science, Wako, Saitama, Japan}
\affiliation{
Institute of Theoretical Physics, Chinese Academy of Sciences, Beijing 100190, China}

\author{Xian-Rong~Zhou (周先荣)} \email{xrzhou@phy.ecnu.edu.cn}
\affiliation{
Department of Physics, East China Normal University,
Shanghai 200241, China}

\date{\today}

\begin{abstract}
We study the charge symmetry breaking (CSB) effect in  mirror hypernuclei using the deformed Skyrme-Hartree-Fock (DSHF)+Bardeen-Cooper-Schrieffer (BCS) model together with the CSB term and pairing interaction. Our model provides a good account for the observations of CSB effect in mirror hypernuclei in the mass region of $A=7\sim16$.We investigate the effect of deformation on the single-$\Lambda$ binding energy differences and we found that, in mirror hypernuclei with mass numbers $A=8$ and $A=9$, deformation has a noticeable impact on the energy difference. We also predict the single-$\Lambda$ binding energy differences in medium-heavy hypernuclei for further confirmation of the CSB effect.
\end{abstract}

\maketitle
\end{CJK*}

\section{Introduction}
\label{intro}

The hyperon-nucleon interaction is important not only for the development of hypernuclei physics~\cite{gal2016strangeness,hao1993hypernucleus} but also for astrophysical studies~\cite{baldo2000hyperon,hofmann2001application,tolos2020strangeness}. Since the observation of the first hyperfragment in an emulsion exposed to cosmic rays in 1953~\cite{danysz1953delayed}, many studies have been conducted at various experimental facilities~\cite{aoki2009nuclear,karpenko2019lambda,ahn2013double,gal2016strangeness,yoshimoto2021first}. Current investigations not only concentrate on $\Lambda$ hypernuclei ($S=-1$) but also focus on $\Lambda\Lambda$ and $\Xi^-$ hypernuclei ($S=-2$). In particular, experimental data on $\Lambda$ hypernuclei are the most abundant.\par 

With the development of advanced experimental facilities, high-resolution measurements of single-$\Lambda$ binding energies for $\Lambda$ hypernuclei have been achieved across a wide mass range from light to heavy hypernuclei. These experimental advancements not only enhance our understanding of $\Lambda$ hypernuclei properties compared to previous studies but also pose challenges to the development of theoretical approaches in hypernuclei structure. To accurately describe the essential features of hypernuclei systems, various forms of $\Lambda N$ interactions have been proposed and discussed, such as the Skyrme types ~\cite{rayet1976self,rayet1981skyrme,millener1988lambda,fernandez1989skyrme,yamamoto2010g,guleria2012study,schulze2014skyrme,schulze2019skyrme}, relativistic types~\cite{marevs1994relativistic,ma1996hypernuclei,tanimura2012description,xu2012single,wang2013new,rong2021new}, the Nijmegen soft-core (NSC) types~\cite{maessen1989soft,rijken1999soft,stoks1999soft}, the Nijmegen Extended-Soft-Core (ESC)  types~\cite{rijken2010status,rijken2010baryon,yamamoto2010hypernuclear,rijken2013baryon,nagels2019extended,nagels2020extended}, as well as chiral effective field theory ($\chi \mathrm{EFT}$)~\cite{polinder2006hyperon,haidenbauer2013hyperon,haidenbauer2020hyperon,song2022test,haidenbauer2023hyperon}. With these effective $\Lambda N$ interactions, extensive studies of $\Lambda $ hypernuclei have been conducted within the framework of various nuclear models, such as the mean field models~\cite{tanimura2012description,xu2012single,wang2013new,rong2021new,lanskoy1998double,tanimura2019clusterization,li2018structure,mei2016generator,mei2018disappearance,Xue22,Xue23,Xue24,Guo22,CFChen22,YFChen22}, the shell models ~\cite{nemura2002ab,le2020jacobi,myo2023structure}, and the antisymmetrized molecular dynamics models~\cite{isaka2011deformation,isaka2015impurity,isaka2016competing,isaka2017effects,isaka2020low}.\par 

The isospin breaking in nuclei comes mainly from the Coulomb interaction.  It  has been known also that  mall isospin symmetry breaking (ISB) effects are recognized in  strong interaction. Mass differences between charged and neutral pions, and differences in short-range interactions~\cite{machleidt2001charge} will cause ISB interactions~\cite{miller2006charge}. The role of ISB interaction has attracted attention in studies such as the isobaric analog states~\cite{suzuki1993charge,colo1998widths,kaneko2014isospin,roca2018nuclear}, the isobaric multiplet mass equation~\cite{gallant2014breakdown,bennett2016isobaric,dong2018generalized}, the Okamoto-Nolen-Schiffer anomaly in the mass differences of mirror nuclei~\cite{brown2000displacement,kaneko2010isospin,dong2018generalized,dong2019effective}, and superallowed Fermi $\beta$ decay~\cite{towner2008improved,towner2010comparative,satula2011microscopic,kaneko2017isospin}. The charge symmetry breaking (CSB) interaction, which arises from $\rho$-$\omega$ and $\pi$-$\eta$ meson couplings in the meson-exchange picture~\cite{miller1990charge,niskanen2002charge}, is a dominant component of ISB. In quantum chromodynamics (QCD), the CSB effect is caused by the mass difference between u$-$ and d$-$ quarks and the partial restoration of the spontaneous symmetry breaking (SSB) in nuclear medium~\cite{miller2006charge}. The importance of the CSB interaction in the nuclear structure of $^{8}$Be has also been explored by using Green's function Monte Carlo calculations~\cite{wiringa2013charge}. Although the CSB interaction in the nucleon-nucleon ($NN$) system has been extensively studied~\cite{muther1999isospin}, its importance in the hyperon-nucleon ($\la N$) interaction is still not well understood and needs further intensive studies.\par 
Significant progress has been achieved in the charge symmetry breaking (CSB) effect in hypernuclei through both experimental observations and theoretical research. Firstly, the experiments by Gajewski et al. and Bohm et al. give 
$B_{\Lambda}(^{4}_{\Lambda}\text{He}) - B_{\Lambda}(^{4}_{\Lambda}\text{H}) = +0.34 \pm 0.08\text{MeV}$, as evidence of CSB effect in the $\Lambda N$ interaction. Then, the mirror hypernuclei with mass number $A=7$ and $A=8$ were also found experimentally~\cite{nakamura2013observation,gogami2016spectroscopy,jurivc1973new}. The updated binding energy difference for $A=7$ hypernuclei ($\Delta B_\Lambda = 0.39\,\text{MeV}$ between $^7_\Lambda\text{He}$ and $^7_\Lambda\text{Be}$) is slightly larger than the $0.34\,\text{MeV}$ found in $A=4$ systems. Furthermore, a set of mirror hypernuclei with mass number $A = 8$ was also found in this experiment, with binding energies of $6.80 \pm0.03(\text{stat})\pm0.04(\text{syst})$~MeV for $^{8}_\Lambda$Li and $6.84\pm0.05(\text{stat})\pm0.04(\text{syst})$~MeV for $^{8}_\Lambda$Be~\cite{jurivc1973new}.
On top of these experimental results, there are some experimental results for the mirror hypernuclei ($^{10}_{\Lambda}$Be,$^{10}_{\Lambda}$B)~\cite{jurivc1973new,cantwell1974binding,gogami2016high}. 
Heavier $\Lambda$ hypernuclei with mass number $A = 12$ and $A = 16$ were also studied in JLab and new experimental data from the $(e,e'K^+)$ reaction was reported~\cite{tang2014experiments,dluzewski1988binding,davis200550,gogami2016high,cusanno2009high,agnello2011hypernuclear}.\par

Theoretically, CSB in hypernuclei has been studied by using several frameworks. Earlier, an important role of the $\Lambda$-$\Sigma^{0}$ mixing related to $\Lambda N$-$\Sigma N$ coupling was pointed out to understand the CSB effect~\cite{dalitz1964electromagnetic}; thus, many models including the $\Lambda$-$\Sigma^{0}$ coupling have been extended to study  the mirror hypernuclei $^{4}_{\Lambda}$(H,He)~\cite{nogga2002hypernuclei,haidenbauer2007hyperon,nogga2013light,gal2015charge,gazda2016charge,schafer2022medium}. More recently, one realistic baryon-baryon potential derived from chiral effective field theory (EFT) was employed, which allow a systematic inclusion of CSB terms as well as $\Lambda$--$\Sigma^0$ mixing contributions~\cite{maessen1989soft,rijken1999soft}. Shell-model calculations incorporating such interactions have successfully reproduced the found $\Delta B_\Lambda$ values in $A=4$ mirror hypernuclei~\cite{gazda2016charge}. In addition, the relativistic mean-field (RMF) model has been extended to study CSB effect in hypernuclei include both hyperon-nucleon interactions and CSB corrections~\cite{sun2025charge}, the extended RMF model reproduces the CSB energy anomaly in some mirror hypernuclei ($A=7$), predicts larger CSB effects in hypernuclei with larger isospins ($A=40$ with $T=5/2$ and $A=48$), and awaits future experiments to confirm the effect in medium-heavy hypernuclei~\cite{sun2025charge}. These methods are particularly useful to investigate the systematics of CSB effects in hypernuclei while ab initio methods are still computationally challenging. Altogether, these theoretical efforts underline the importance of CSB in testing fundamental symmetries and understanding the structure of hypernuclei matter.\par  

However, these approaches face challenges in consistently describing both light and heavy hypernuclei within a unified framework except for the RMF models in which deformation effects was not considered~\cite{sun2025charge}. The Skyrme-Hartree-Fock (SHF) model is known as a powerful approach not only for normal nuclei but also for hypernuclei based on the density functional theory (DFT), which can be globally applied from light to heavy nuclei and hypernuclei based on universal energy density functionals (EDFs)~\cite{vautherin1972hartree,Vautherin73,bender2003self,Cugnon00,Zhou07}. It is known that $^{12}\mathrm{C}$ is an axially deformed oblate nucleus~\cite{jin2020study,Suzuki03,sagawa2004deformations,Schulze10,cui2015shape,cui2017investigation,cui2017beyond},
and also the shrinkage effect of $^{7}_{\Lambda}$Li may be associated with the deformation of the core nucleus~\cite{Tanida2001}.
In this work, we present a systematic investigation of the CSB interaction of hypernuclei across the mass range $A=7\sim40$ using the deformed SHF (DSHF) +BCS approach accommodating the quadrupole deformation degree of freedom and the pairing correlations.\par

The paper is organized as follows. In Sec.~\ref{sec2}, the theoretical DSHF+BCS approach for hypernuclei and the $\Lambda N$ CSB interaction are introduced. In Sec.~\ref{sec3}, calculated results and discussions for the mirror hypernuclei ($^{7}_{\Lambda}$He, $^{7}_{\Lambda}$Be), 
($^{8}_{\Lambda}$Li, $^{8}_{\Lambda}$Be), 
($^{9}_{\Lambda}$Li, $^{9}_{\Lambda}$B), 
($^{10}_{\Lambda}$Be, $^{10}_{\Lambda}$B), 
($^{12}_{\Lambda}$B, $^{12}_{\Lambda}$C), 
($^{16}_{\Lambda}$N, $^{16}_{\Lambda}$O), 
($^{32}_{\Lambda}$P, $^{32}_{\Lambda}$S), and ($^{40}_{\Lambda}$Ca, $^{40}_{\Lambda}$K) are presented. Finally, a summary is given in Sec.~\ref{sec4}.\par


\section{Theoretical framework}
\label{sec2}

\subsection{DSHF approach}

In the DSHF approach, the total energy of a hypernuclei is given by~\cite{Cugnon00}
\be\label{e:Energy}
 E = \int d^3\rv\; \eps(\rv) \:,
\ee
where the energy-density functional is
\be
 \eps = \eps_N[\rho_n,\rho_p,\tau_n,\tau_p,\Jv_n,\Jv_p] +
        \eps_\la[\rho_n,\rho_p,\rol,\tau_\la,\Jv_N,\Jv_\la] \:,
\ee
with $\eps_N$ and $\eps_\la$ as the contributions from $NN$ and $\la N$ interactions together with the kinetic energies of nucleons and hyperon, respectively. For the nucleonic functional $\eps_N$, we use the Skyrme forces SLy4 or SLy5~\cite{Chabanat98}, which will be discussed in detail later. The one-body density $\rho_q$,
kinetic density $\tau_q$, and spin-orbit current density $\Jv_q$ read
\be
 \Big[ \rho_q,\; \tau_q,\; \Jv_q \Big] =
  \sum_{k=1}^{N_q} {n_q^k} \Big[
  |\phi_q^k|^2 ,\;
  |\Nv\phi_q^k|^2 ,\;
  {\phi_q^k}^* (\Nv\phi_q^k \times \bm{\sigma})/i
 \Big] \:,
\label{e:rho}
\ee
where,$\phi_q^k$ $(k=1,\cdots,N_q)$ denote the single-particle (s.p.) wave functions corresponding to the $k$-th occupied states for distinct particles $q=n,p,\Lambda$. The occupation probabilities $n_q^k$ are computed by incorporating pairing correlations within the BCS approximation, with this treatment applied exclusively to nucleons. The nucleon-nucleon pairing interaction is modeled as a density-dependent surface-type force~\cite{Tajima93,Bender99},
\be\label{e:pair}
 V_q\left(\rv_1, \rv_2\right) = V_q'
 \left[ 1-\frac{\ron((\rv_1+\rv_2)/2)}{0.16~\fmq} \right]
 \delta\left(\rv_1-\rv_2\right) \:.
\ee
In this framework, pairing strengths are set to $V'_p = V'_n = -410\;\text{MeVfm}^3$ for light-mass nuclei~\cite{Suzuki03}, whereas values of $V'_p = -1146\;\text{MeVfm}^3$ and $V'_n = -999\;\text{MeVfm}^3$ are adopted for medium-mass and heavy-mass nuclei~\cite{Zhou07}. BCS calculations implement a smooth energy cutoff for the pairing window~\cite{Bender00}. For systems with an odd number of nucleons, the orbit occupied by the unpaired nucleon is blocked following the procedure described in Ref.~\cite{Ring80}.
Through the variation of the total energy Eq.~(\ref{e:Energy}), one derives the SHF Sch\"odinger equation for both nucleons and hyperons~\cite{Cugnon00},
\be
 \Big[ - \Nv \cdot \frac{1}{2m_q^*(\rv)} \Nv
 + V_q(\rv) -i\Wv_q(\rv) \cdot (\Nv\times\bm{\sigma})
 \Big] \phi_q^k(\rv) = e_q^k \phi_q^k(\rv) \:.
\label{e:se}
\ee
In this equation, $V_q(\rv)$ represents the central component of the mean-field potential, which depends on the densities, while $\Wv_q(\rv)$ corresponds to the spin-orbit potential~\cite{Vautherin73,Bender99}.\par 
For the Skyrme-type $\Lambda N$ interactions, $\eps_\la$ is given as~\cite{schulze2014skyrme}
\bal\label{e:ef}
 \eps_\la =& \frac{\tau_\la}{2m_\la} + a_0 \rol\ron
  + a_1 \left(\rol \tau_N + \ron \tau_\la \right)\notag\\&
  - a_2 \left(\rol \Delta\ron + \ron \Delta\rol \right)/2  \\&
  + a_3 \rol\ron^{1+\alpha}- a_4 \left(\rol\Nv\cdot\Jv_N + \ron\Nv\cdot\Jv_\la \right) \notag
\:.
\eal
Since the previous studies have shown this spin-orbit splitting to be extremely small (about 0.1 MeV)~\cite{Xue22}, it is neglected in the present work. The $a_3$ term describes a many-body interaction originating from higher-order $G$-matrices~\cite{millener1988lambda,Lanskoy97,guleria2012study}. We utilize the Skyrme force parameter set SLL4~\cite{schulze2014skyrme,schulze2019skyrme} for the $\Lambda N$ interaction, which was optimized to fit the complete existing dataset of single-$\Lambda$ hypernuclei in spherical SHF calculations~\cite{schulze2019skyrme}. For clarity, we reiterate the parameters here~\cite{schulze2019skyrme}: $\alpha = 1$, $a_{0,1,2,3}$ = [-322.0{\rm MeVfm$^3$}, 15.75{\rm MeVfm$^5$}, 19.63{\rm MeVfm$^5$}, 715.0{\rm MeVfm$^{3(1+\alpha)}$}].\par

Then one obtains the corresponding SHF mean field for a hyperon, 
\begin{equation}
	\begin{aligned}
		V_{\Lambda}= & a_0 \rho_N+a_1 \tau_N-a_2 \Delta \rho_N+a_3 \rho_N^{1+\alpha},
	\end{aligned}
\label{VL}
\end{equation}
and the $\Lambda N$ interaction contributions for nucleonic mean fields are
\begin{equation}
	\begin{aligned}
		V_N^{(\Lambda)}= & a_0 \rho_{\Lambda}+a_1 \tau_{\Lambda}-a_2 \Delta \rho_{\Lambda}
        +a_3(1+\alpha) \rho_{\Lambda} \rho_N^\alpha.
	\end{aligned}
\label{Vq}
\end{equation}
\par 
In the present calculations, the deformed SHF Schr\"odinger equation is solved in the cylindrical coordinates $(r,z)$, under the assumption of axial symmetry of the mean fields.
The optimal quadrupole deformation parameters are defined as
\be
 \beta_2^{(q)} = \sqrt{\frac{\pi}{5}}
 \frac{\langle 2z^2-r^2 \rangle_q}{\langle z^2+r^2 \rangle_q}.
\ee

\subsection{$\Lambda N$ CSB interaction}

We use a $\Lambda N$ CSB interaction as in Ref.~\cite{sun2025charge} in the contact form as
\be  \label{LNCSB}
V_{\Lambda N}^{\text{CSB}} = -\frac{v_0}{2}\tau_z^N\delta(\mathbf{r}_\Lambda - \mathbf{r}_N),
\ee
where $\tau_z^N$ denotes the third component of isospin $+1$ for neutrons and $-1$ for protons. This form is a straightforward extension of $NN$ CSB interaction~\cite{sagawa1995isospin}, which depends on $(\tau_z^1+\tau_z^2)$. Since a hyperon has no isospin, one of the isospin operators is dropped in Eq.~(\ref{LNCSB}) for $\Lambda N$ sector. The CSB interaction is proportional to the strength parameter $v_0$. The EDF from the $\Lambda N$ CSB channel provides an  additional term, 
\be
\eps_{\Lambda N}^{\text{CSB}} = -\frac{\textcolor{black}{v_0}}{2}\rho_{\Lambda}(\rho_{n}-\rho_{p}).
\label{CSB_energy}
\ee
For the $\Lambda$ energy term, the corresponding CSB mean field expression is
\begin{equation}
	\begin{aligned}
 V_{\Lambda}^{\text{CSB}}=-\frac{\textcolor{black}{v_0}}2(\rho_n-\rho_p),
 \label{CSB_lambda}
	\end{aligned}
\end{equation}
and the neutron and proton energy terms, the corresponding CSB mean field expressions are
\begin{equation}
	\begin{aligned}
   V_{n}^{\text{CSB}}=-\frac{\textcolor{black}{v_0}}2\rho_\Lambda,
	\end{aligned}
    \label{CSB_neutron}
\end{equation}
\begin{equation}
	\begin{aligned}
    V_{p}^{\text{CSB}}=+\frac{\textcolor{black}{v_0}}2\rho_\Lambda.
      \label{CSB_proton}
	\end{aligned}
\end{equation}
For a positive value of $v_0$, the $\Lambda N$ CSB interaction gives attractive contributions for a hyperon and neutron potential, whereas it is repulsive for protons. To account for the $\Lambda N$ CSB interaction, Eqs.~(\ref{CSB_energy})--(\ref{CSB_proton}) should be incorporated into Eqs.~(\ref{e:ef})--(\ref{Vq}) in the SHF calculation.\par 
The single-$\Lambda$ binding energy $B_{\Lambda}$ of a $\Lambda$ hypernuclei is given by 
\begin{equation}
	\begin{aligned}
         B_{\Lambda}(Z, N, N_\Lambda=1) = B(^A_\Lambda Z) - B(^{A-1}Z).
	\end{aligned}
\end{equation}
The binding energies of the $\Lambda$ hypernuclei $^A_\Lambda Z$ and its corresponding core nucleus $^{A-1}Z$ are denoted by $B(^A_\Lambda Z)$ and $B(^{A-1}Z)$, respectively. The difference in single-$\Lambda$ binding energies $B_{\Lambda}$ between mirror hypernuclei for $N \geq Z$ is calculated as follows: 
\begin{equation}
	\begin{aligned}
        \Delta B_{\Lambda}(A = N + Z+1) = B_{\Lambda}(Z, N, 1) - B_{\Lambda}(N, Z, 1).
	\end{aligned}
\end{equation}
\par

\section{Results and discussion}
\label{sec3}

\begin{figure}[!ht]
 \begin{center}
  \includegraphics[width=0.95\linewidth]{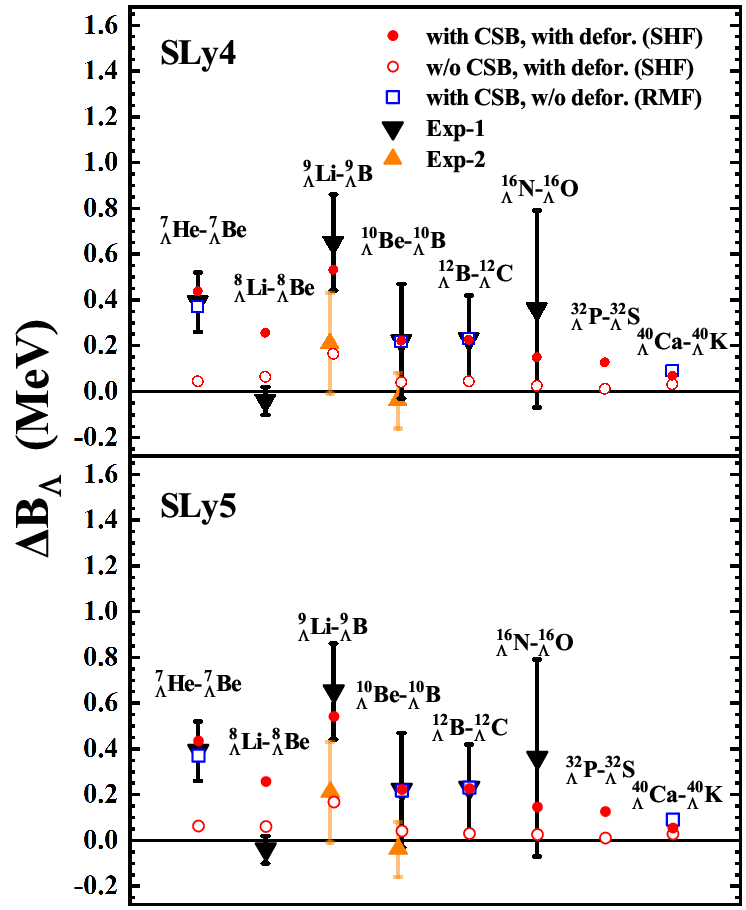}
  \vskip-3mm
\caption{
The differences $\Delta B_{\Lambda}$ of the single-$\Lambda$ binding energies between the mirror hypernuclei 
($^{7}_{\Lambda}$He, $^{7}_{\Lambda}$Be), 
($^{8}_{\Lambda}$Li, $^{8}_{\Lambda}$Be), 
($^{9}_{\Lambda}$Li, $^{9}_{\Lambda}$B), 
($^{10}_{\Lambda}$Be, $^{10}_{\Lambda}$B), 
($^{12}_{\Lambda}$B, $^{12}_{\Lambda}$C), 
($^{16}_{\Lambda}$N, $^{16}_{\Lambda}$O), 
($^{32}_{\Lambda}$P, $^{32}_{\Lambda}$S), and ($^{40}_{\Lambda}$Ca, $^{40}_{\Lambda}$K) obtained by the DSHF+BCS approach with and without the $\Lambda N$ CSB interaction. The red filled (open) circles  represent the calculated results with (without) $\Lambda N$ CSB interaction in the DSHF+BCS approach, while blue open squares correspond to the binding energy differences of four pairs of mirror hypernuclei calculated by the RMF model~\cite{sun2025charge} with CSB interaction, respectively. 
The upper (lower) panel shows results with $NN$ interaction SLy4 (SLy5). 
The experimental data are denoted by black inverted triangles, with vertical black lines indicating the experimental uncertainties (Exp-1). Data are from Refs.~\cite{botta2017binding,gogami2016high,hasegawa1996spectroscopic,tang2014experiments,davis200550,pile1991study}. 
In the cases of mirror hypernuclei with $A = 9$~\cite{jurivc1973new,davis200550} and $A = 10$~\cite{cantwell1974binding,gogami2016high,davis200550}, two sets of experimental data are available; the second set is represented by orange upright triangles and orange error bars (Exp-2). }
\label{plt.Pjsp}
 \end{center}
\end{figure}


\begin{table*}
\renewcommand{\arraystretch}{1.3}  
\caption{We performed calculations using the DSHF+BCS approach.The single-$\Lambda$ binding energies $B_{\Lambda}$ as well as the difference $\Delta B_{\Lambda(\text{SHF})}$ in mirror hypernuclei calculated using $NN$ interaction SLy4 and the $\Lambda N$ SLL4 interaction with or without the $\Lambda N$ CSB interaction are presented. Results by RMF model~\cite{sun2025charge}, by no-core shell model (\text{NLO13, NLO19})~\cite{le2023ab}, by Gal~\cite{gal2015charge} and by Hiyama et al.~\cite{hiyama2009} are included for comparison. The quadrupole deformation parameter $\beta_2$ is also listed. For comparison, available experimental data are listed with statistical uncertainties. All energies are in units of MeV.}
\label{sly4}
\centering
\begin{tabular}{cccccccccccc}
\toprule
\multirow{2}{*}{} 
& \multirow{2}{*}{}  
& \multicolumn{4}{c}{\textbf{w/o CSB}} 
& \multicolumn{4}{c}{\textbf{with CSB}}
& \multirow{2}{*}{$\Delta B_{\Lambda(\text{Expt.})}$}
& \multirow{2}{*}{$B_{\Lambda(\text{Expt.})}$}  \\ 
\cmidrule(lr){3-6} \cmidrule(lr){7-10}
 & &  $\beta_2$ & $B_{\Lambda}$ & $\Delta B_{\Lambda(\text{SHF})}$ &$\Delta B_{\Lambda(\text{RMF})}$& $\beta_2$ & $B_{\Lambda}$ & $\Delta B_{\Lambda(\text{SHF})}$ &$\Delta B_{\Lambda}$&& \\ 
\midrule
\multirow{2}{*}{$A=7$} 
& $_{\Lambda}^{7}$He   & 0.017  & 5.721  & \multirow{2}{*}{0.044} &\multirow{2}{*}{0.003} &  0.009  & 5.927   & \multirow{2}{*}{0.436} &\multirow{2}{*}{0.368} & \multirow{2}{*}{$0.39\pm0.13$}& $5.55\pm0.10$~\cite{gogami2016spectroscopy}  \\
& $_{\Lambda}^{7}$Be  & 0.001  & 5.677   &  & & 0.001  & 5.492   & &(\text{RMF})~\cite{sun2025charge} && $5.16\pm0.08$~\cite{jurivc1973new} \\

\midrule
\multirow{8}{*}{$A=8$} 
& \multirow{4}{*}{$_{\Lambda}^{8}$Li}    & \multirow{4}{*}{0.283}  & \multirow{4}{*}{6.761}  & \multirow{8}{*}{0.063} &\multirow{8}{*}{-}  & \multirow{4}{*}{0.297}  & \multirow{4}{*}{6.866}   & \multirow{8}{*}{0.256} &\multirow{2}{*}{-0.178} & \multirow{8}{*}{$-0.04\pm0.06$} & \multirow{4}{*}{$6.80\pm0.03$~\cite{jurivc1973new}}\\
& &  &   &  &  &  & & &(\text{NLO13})~\cite{le2023ab} &&  \\
& &  &   &  &  &  & & &\multirow{2}{*}{-0.143} &&  \\
& &  &   &  &  &  & & &(\text{NLO19})~\cite{le2023ab} &&  \\
& \multirow{4}{*}{$_{\Lambda}^{8}$Be}    & \multirow{4}{*}{0.321}  & \multirow{4}{*}{6.698} &  &  & \multirow{4}{*}{0.324}  & \multirow{4}{*}{6.610} &  &\multirow{2}{*}{-0.160} && \multirow{4}{*}{$6.84\pm0.05$~\cite{jurivc1973new}} \\
& &  &   &  &  &  & & &(\text{Hiyama})~\cite{hiyama2009} &&  \\
& &  &   &  &  &  & & &\multirow{2}{*}{-0.049} &&  \\
& &  &   &  &  &  & & &(\text{Gal})~\cite{gal2015charge} &&  \\

\midrule
\multirow{4}{*}{$A=9$} 
& \multirow{2}{*}{$_{\Lambda}^{9}$Li}
   & \multirow{2}{*}{$-$0.167}  & \multirow{2}{*}{7.937}  & \multirow{4}{*}{0.164} &\multirow{4}{*}{-} & \multirow{2}{*}{$-$0.164}  & \multirow{2}{*}{8.132}   & \multirow{4}{*}{0.530 }  &\multirow{4}{*}{0.054} & \multirow{2}{*}{$0.65\pm0.21$} & $8.53\pm0.15$~\cite{jurivc1973new} \\
& &  &   &  &  &  & & & &~\cite{jurivc1973new}& $8.50\pm0.12$~\cite{davis200550} \\

& \multirow{2}{*}{$_{\Lambda}^{9}$B}  
  &\multirow{2}{*}{0.282}  & \multirow{2}{*}{7.773}  &&  & \multirow{2}{*}{0.284}   & \multirow{2}{*}{7.602}   & & (\text{Gal})~\cite{gal2015charge} &\multirow{2}{*}{$0.21\pm0.22$}& $7.88\pm0.15$~\cite{jurivc1973new} \\
  
&  &  &   &  &  &  & & && ~\cite{davis200550}& $8.29\pm0.18$~\cite{davis200550}\\

\midrule
\multirow{4}{*}{$A=10$} 
& \multirow{2}{*}{$_{\Lambda}^{10}$Be}   & \multirow{2}{*}{0.265}  & \multirow{2}{*}{8.906}  & \multirow{4}{*}{0.040} & \multirow{4}{*}{0.015} &  \multirow{2}{*}{0.278}  & \multirow{2}{*}{9.006}   & \multirow{4}{*}{0.220} & \multirow{2}{*}{0.217} & \multirow{2}{*}{$0.22\pm0.25$}& $9.11\pm0.22$~\cite{cantwell1974binding}  \\

&   &  &   &  &  &  & & &(\text{RMF})~\cite{sun2025charge} &~\cite{cantwell1974binding,davis200550} & $8.60\pm0.07$~\cite{gogami2016high}\\

& \multirow{2}{*}{$_{\Lambda}^{10}$B}  & \multirow{2}{*}{0.278}  & \multirow{2}{*}{8.866} & & & \multirow{2}{*}{0.279}   & \multirow{2}{*}{8.786}  & &  \multirow{2}{*}{0.136} &\multirow{2}{*}{$-0.04\pm0.12$}& $8.89\pm0.12$~\cite{davis200550}  \\
  
&    &  &   &  &  &  & & &(\text{Gal})~\cite{gal2015charge}&~\cite{gogami2016high}& $8.64\pm0.1$~\cite{gogami2016high}  \\
       
\midrule
\multirow{2}{*}{$A=12$} 
& $_{\Lambda}^{12}$B   & 0.103  & 10.969  & \multirow{2}{*}{0.029}& \multirow{2}{*}{0.021} & 0.109  & 11.088   & \multirow{2}{*}{0.225} & \multirow{2}{*}{0.229}& \multirow{2}{*}{$0.23\pm0.19$}  & $11.529\pm0.025$~\cite{tang2014experiments}\\
& $_{\Lambda}^{12}$C  & 0.120  & 10.940   &  &
       & 0.113  & 10.863   &  & (\text{RMF})~\cite{sun2025charge} & & $11.30\pm0.19$~\cite{dluzewski1988binding, davis200550, gogami2016high}\\
       
\midrule
\multirow{2}{*}{$A=16$} 
& $_{\Lambda}^{16}$N    & \multirow{2}{*}{0}  & 13.630  & \multirow{2}{*}{0.024} &\multirow{2}{*}{-} & \multirow{2}{*}{0}  & 13.707   & \multirow{2}{*}{0.148} & \multirow{2}{*}{-}&  \multirow{2}{*}{$0.36\pm0.43$}& $13.76\pm0.16$~\cite{cusanno2009high} \\
& $_{\Lambda}^{16}$O  &   & 13.606 &  & 
&   & 13.559   &     &&& $13.4\pm0.4$~\cite{agnello2011hypernuclear}  \\

\midrule
\multirow{2}{*}{$A=32$} 
& $_{\Lambda}^{32}$P   & \multirow{2}{*}{0}  & 18.747  & \multirow{2}{*}{0.010} &\multirow{2}{*}{-} & \multirow{2}{*}{0}  & 18.832   & \multirow{2}{*}{0.125} & \multirow{2}{*}{-} &  \multirow{2}{*}{-} & - \\
& $_{\Lambda}^{32}$S  &   & 18.737  & & 
       &   & 18.707  &  &                      && -  \\       
       
\midrule
\multirow{2}{*}{$A=40$}
& $_{\Lambda}^{40}$K   & 0.019  & 20.022   & \multirow{2}{*}{0.031} 
& \multirow{2}{*}{0.024}         & 0.019   & 20.079                 & \multirow{2}{*}{0.067} & \multirow{2}{*}{0.089} & \multirow{2}{*}{-}  & - \\
& $_{\Lambda}^{40}$Ca  & 0.019  & 19.991  & & & 0.025  & 20.012    & &(\text{RMF})~\cite{sun2025charge} &  & $18.7\pm1.1$~\cite{pile1991study} \\

\bottomrule
\end{tabular}
\end{table*}

\begin{table*}
\renewcommand{\arraystretch}{1.3}
\caption{The same as Table~\ref{sly4}, but with the SLy5 $NN$ interaction.}
\label{sly5}
\centering
\begin{tabular}{cccccccccccc}
\toprule
\multirow{2}{*}{} & \multirow{2}{*}{}  & \multicolumn{4}{c}{\textbf{w/o CSB}} & \multicolumn{4}{c}{\textbf{with CSB}} & \multirow{2}{*}{$\Delta B_{\Lambda(\text{Expt.})}$} & \multirow{2}{*}{$B_{\Lambda(\text{Expt.})}$}\\ 
\cmidrule(lr){3-6} \cmidrule(lr){7-10}
 &  & $\beta_2$ & $B_{\Lambda}$ & $\Delta B_{\Lambda(\text{SHF})}$ &$\Delta B_{\Lambda(\text{RMF})}$& $\beta_2$ & $B_{\Lambda}$ & $\Delta B_{\Lambda(\text{SHF})}$ &$\Delta B_{\Lambda}$& &
\\ 
\midrule
\multirow{2}{*}{$A=7$} 
& $_{\Lambda}^{7}$He  
 & 0.017  & 5.728  & \multirow{2}{*}{0.061} &\multirow{2}{*}{0.003}& 0.017  & 5.895   & \multirow{2}{*}{0.431}&\multirow{2}{*}{0.368} & \multirow{2}{*}{$0.39\pm0.13$} & $5.55\pm0.10$~\cite{gogami2016spectroscopy}\\
& $_{\Lambda}^{7}$Be     & 0.001  & 5.667 &  &  & 0.001  & 5.464  & &(\text{RMF})~\cite{sun2025charge} & & $5.16\pm0.08$~\cite{jurivc1973new} \\

\midrule
\multirow{8}{*}{$A=8$} 
& \multirow{4}{*}{$_{\Lambda}^{8}$Li} 
  & \multirow{4}{*}{0.309}  & \multirow{4}{*}{6.759}  & \multirow{8}{*}{0.059} &\multirow{8}{*}{-} & \multirow{4}{*}{0.284}  & \multirow{4}{*}{6.685}   & \multirow{8}{*}{0.254} &\multirow{2}{*}{-0.178}  & \multirow{8}{*}{$-0.04\pm0.06$}  & \multirow{4}{*}{$6.80\pm0.03$~\cite{jurivc1973new}}\\
& &  &   &  &  &  & & &(\text{NLO13})~\cite{le2023ab} &&  \\
& &  &   &  &  &  & & &\multirow{2}{*}{-0.143} &&  \\
& &  &   &  &  &  & & &(\text{NLO19})~\cite{le2023ab} &&  \\
& \multirow{4}{*}{$_{\Lambda}^{8}$Be}    & \multirow{4}{*}{0.321}  & \multirow{4}{*}{6.700}  & &  & \multirow{4}{*}{0.309}  & \multirow{4}{*}{6.431}   &&\multirow{2}{*}{-0.160} & & \multirow{4}{*}{$6.84\pm0.05$~\cite{jurivc1973new}}\\
& &  &   &  &  &  & & &(\text{Hiyama})~\cite{hiyama2009} &&  \\
& &  &   &  &  &  & & &\multirow{2}{*}{-0.049} &&  \\
& &  &   &  &  &  & & &(\text{Gal})~\cite{gal2015charge} &&  \\
 
\midrule
\multirow{4}{*}{$A=9$} 
& \multirow{2}{*}{$_{\Lambda}^{9}$Li}
    & \multirow{2}{*}{$-$0.167}  & \multirow{2}{*}{7.924}
 & \multirow{4}{*}{0.166}&\multirow{4}{*}{-} & \multirow{2}{*}{$-$0.167}  & \multirow{2}{*}{8.031}   & \multirow{4}{*}{0.539 } &\multirow{4}{*}{0.054}  & \multirow{2}{*}{$0.65\pm0.21$} & $8.53\pm0.15$~\cite{jurivc1973new}\\
& &  &   &  &  &  &&& &~\cite{jurivc1973new} & $8.50\pm0.12$~\cite{davis200550} \\

& \multirow{2}{*}{$_{\Lambda}^{9}$B}  
  &\multirow{2}{*}{0.295}  & \multirow{2}{*}{7.758} & & & \multirow{2}{*}{0.282}   & \multirow{2}{*}{7.492} &  & (\text{Gal})~\cite{gal2015charge} &\multirow{2}{*}{$0.21\pm0.22$} & $7.88\pm0.15$~\cite{jurivc1973new}\\
  
&   &  &   &  &  & & & && ~\cite{davis200550}& $8.29\pm0.18$~\cite{davis200550} \\

\midrule
\multirow{4}{*}{$A=10$} 
& \multirow{2}{*}{$_{\Lambda}^{10}$Be}
    & \multirow{2}{*}{0.280}  & \multirow{2}{*}{8.900}  & \multirow{4}{*}{0.039}& \multirow{4}{*}{0.015} & \multirow{2}{*}{0.266}  & \multirow{2}{*}{8.786}   & \multirow{4}{*}{0.220 } & \multirow{2}{*}{0.217}& \multirow{2}{*}{$0.22\pm0.25$} & $9.11\pm0.22$~\cite{cantwell1974binding}\\

&                                   &&&  &   &  &  &  & (\text{RMF})~\cite{sun2025charge}&~\cite{cantwell1974binding,davis200550}& $8.60\pm0.07$~\cite{gogami2016high} \\

& \multirow{2}{*}{$_{\Lambda}^{10}$B}  
  & \multirow{2}{*}{0.292}  & \multirow{2}{*}{8.861} & & & \multirow{2}{*}{0.279}   & \multirow{2}{*}{8.565}  &  & \multirow{2}{*}{0.136} &\multirow{2}{*}{$-0.04\pm0.12$}&$8.89\pm0.12$~\cite{davis200550} \\
  
&                                &&   &  &   &  &  &  &(\text{Gal})~\cite{gal2015charge}&~\cite{gogami2016high}  & $8.64\pm0.1$~\cite{gogami2016high} \\

\midrule
\multirow{2}{*}{$A=12$} 
& $_{\Lambda}^{12}$B 
  & -0.108  & 10.956  & \multirow{2}{*}{0.027}& \multirow{2}{*}{0.021} & 0.104  & 10.857   & \multirow{2}{*}{0.225}& \multirow{2}{*}{0.229} & \multirow{2}{*}{$0.23\pm0.19$} & $11.529\pm0.025$~\cite{tang2014experiments} \\
& $_{\Lambda}^{12}$C   & 0.119  & 10.929  & & 
       & 0.103  & 10.632   && (\text{RMF})~\cite{sun2025charge}                     & & $11.30\pm0.19$~\cite{dluzewski1988binding,davis200550,gogami2016high} \\
       
\midrule
\multirow{2}{*}{$A=16$} 
& $_{\Lambda}^{16}$N  
  & \multirow{2}{*}{0}  & 13.634  & \multirow{2}{*}{0.024}& \multirow{2}{*}{-} & \multirow{2}{*}{0}  & 13.290   & \multirow{2}{*}{0.144}& \multirow{2}{*}{-} & \multirow{2}{*}{$0.36\pm0.43$}& $13.76\pm0.16$~\cite{cusanno2009high} \\
& $_{\Lambda}^{16}$O   &   & 13.610   & &
       &   & 13.146   &  &                    && $13.4\pm0.4$~\cite{agnello2011hypernuclear} \\

\midrule
\multirow{2}{*}{$A=32$} 
& $_{\Lambda}^{32}$P   & \multirow{2}{*}{0}  & 18.745  & \multirow{2}{*}{0.009}& \multirow{2}{*}{-} & \multirow{2}{*}{0}  & 18.399   & \multirow{2}{*}{0.125} & \multirow{2}{*}{-}& \multirow{2}{*}{-}& \multirow{2}{*}{-} \\
& $_{\Lambda}^{32}$S   &  & 18.736 &  & 
       &   & 18.274 &  &                      &&  \\       
       
\midrule
\multirow{2}{*}{$A=40$}
& $_{\Lambda}^{40}$K    & 0.019  & 20.037   & \multirow{2}{*}{0.027}  & \multirow{2}{*}{0.024}
         & 0.019   & 19.256                 & \multirow{2}{*}{0.051} & \multirow{2}{*}{0.089} & \multirow{2}{*}{-} & - \\
& $_{\Lambda}^{40}$Ca    & 0.019  & 20.010 & && 0.025  & 19.205 &  &(\text{RMF})~\cite{sun2025charge} && $18.7\pm1.1$~\cite{pile1991study} \\

\bottomrule
\end{tabular}
\end{table*}

\begin{figure}[!ht]
 \begin{center}
  \includegraphics[width=0.95\linewidth]{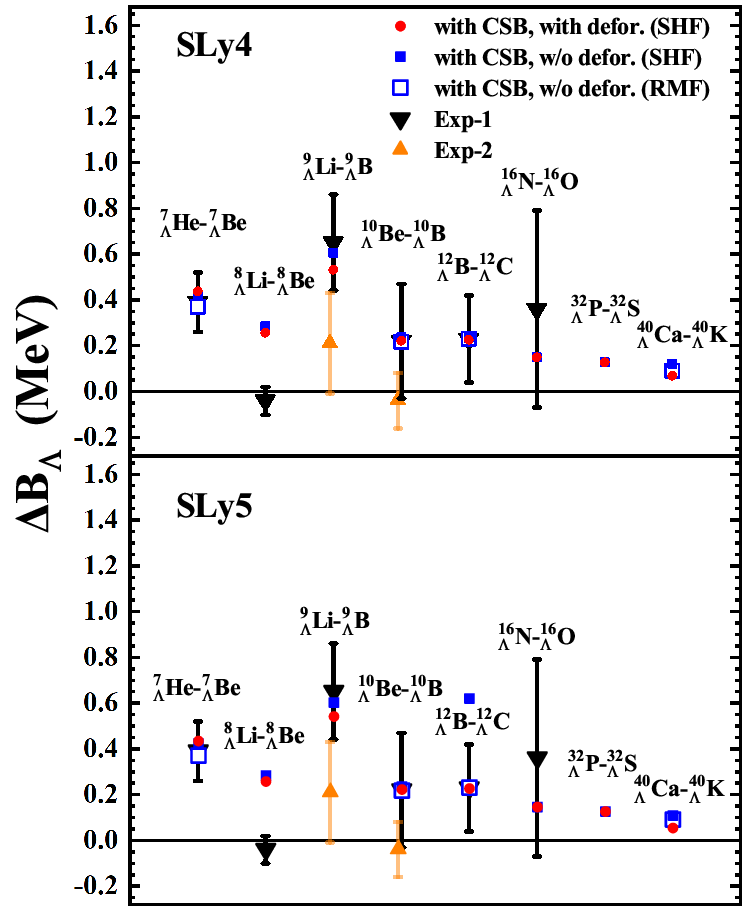}
  \caption{The differences $\Delta B_{\Lambda}$ of the single-$\Lambda$ binding energies between the mirror hypernuclei calculated  by the DSHF+BCS approach including the $\Lambda N$ CSB interaction, with and without nuclear deformation. The calculation results are obtained by using the $NN$ interaction SLy4 and $\Lambda N$ interaction SLL4. The red filled circles (blue filled squares) represent the calculated results with (without) nuclear deformation from the DSHF+BCS approach, while blue open squares correspond to the binding energy differences calculated by the RMF model~\cite{sun2025charge} with CSB interaction, respectively. 
The upper (lower) panel shows results with $NN$ interaction SLy4 (SLy5). 
The experimental data are denoted by black inverted triangles, with vertical black lines indicating the experimental uncertainties (Exp-1). Data are from Refs.~\cite{botta2017binding,gogami2016high,hasegawa1996spectroscopic,tang2014experiments,davis200550,pile1991study}. 
In the cases of mirror hypernuclei with $A = 9$~\cite{jurivc1973new,davis200550} and $A = 10$~\cite{cantwell1974binding,gogami2016high,davis200550}, two sets of experimental data are available; the second set is represented by orange upright triangles and orange error bars (Exp-2).
}
   \label{deformed}
 \end{center}
\end{figure}

\begin{table*}[!ht]
\renewcommand{\arraystretch}{1.3}
\caption{The single-$\Lambda$ binding energies $B_{\Lambda}$ as well as the difference $\Delta B_{\Lambda(\text{SHF})}$ in mirror hypernuclei calculated using Skyrme $NN$ interaction SLy4 and the $\Lambda N$ SLL4 interaction with and without nuclear deformation are presented. Here, $\Delta B_{\Lambda(\text{RMF})}$ refers to the binding energy differences of four pairs of mirror hypernuclei obtained by the RMF model in Ref.~\cite{sun2025charge}. The quadrupole deformation parameter $\beta_2$ is also listed. For comparison, available experimental data are listed with statistical uncertainties. All energies are in units of MeV.}
\label{tab:deformation}
\centering
\begin{tabular}{cccccccccc}
\toprule
\multirow{2}{*}{} & \multirow{2}{*}{}  & \multicolumn{2}{c}{\textbf{w/o deformation}} & \multicolumn{3}{c}{\textbf{with deformation}} & \multirow{2}{*}{$\Delta B_{\Lambda(\text{RMF})}$~\cite{sun2025charge}} & \multirow{2}{*}{$\Delta B_{\Lambda(\text{Expt.})}$}& \multirow{2}{*}{$B_{\Lambda(\text{Expt.})}$} \\ 
\cmidrule(lr){3-4} \cmidrule(lr){5-7}
 & & $B_{\Lambda}$ & $\Delta B_{\Lambda(\text{SHF})}$ & $\beta_2$ & $B_{\Lambda}$ & $\Delta B_{\Lambda(\text{SHF})}$ & &\\  
\midrule     
\multirow{2}{*}{$A=7$} 
& $_{\Lambda}^{7}$He 
 & 5.915  & \multirow{2}{*}{0.423} & 0.009  & 5.927   & \multirow{2}{*}{0.436} & \multirow{2}{*}{0.368}& \multirow{2}{*}{$0.39\pm0.13$} & $5.55\pm0.10$~\cite{gogami2016spectroscopy} \\
& $_{\Lambda}^{7}$Be  & 5.492   &  & 0.001  & 5.492  & & & & $5.16\pm0.08$~\cite{jurivc1973new}  \\

\midrule
\multirow{2}{*}{$A=8$} 
& $_{\Lambda}^{8}$Li  
    & 6.333  & \multirow{2}{*}{0.283} & 0.297  & 6.866   & \multirow{2}{*}{0.256}& \multirow{2}{*}{-} & \multirow{2}{*}{$$-$0.04\pm0.06$}& $6.80\pm0.03$~\cite{jurivc1973new} \\
& $_{\Lambda}^{8}$Be    & 6.050   &  & 0.324  & 6.610 &   & &  & $6.84\pm0.05$~\cite{jurivc1973new}\\

\midrule
\multirow{4}{*}{$A=9$} 
& \multirow{2}{*}{$_{\Lambda}^{9}$Li}
     & \multirow{2}{*}{6.903}  & \multirow{4}{*}{0.604} & \multirow{2}{*}{$-$0.164}  & \multirow{2}{*}{8.132}   & \multirow{4}{*}{0.530 }& \multirow{4}{*}{-} & \multirow{2}{*}{$0.65\pm0.21$} & $8.53\pm0.15$~\cite{jurivc1973new}\\

&                                   
    &   &  &  &  & &&~\cite{jurivc1973new} & $8.50\pm0.12$~\cite{davis200550} \\

& \multirow{2}{*}{$_{\Lambda}^{9}$B}  
    & \multirow{2}{*}{6.299}  & & \multirow{2}{*}{0.284}   & \multirow{2}{*}{7.602}   &  &&\multirow{2}{*}{$0.21\pm0.22$} & $7.88\pm0.15$~\cite{jurivc1973new} \\
  
&                                     &   &  &  &  & &&~\cite{davis200550}& $8.29\pm0.18$~\cite{davis200550} \\

\midrule
\multirow{4}{*}{$A=10$} 
& \multirow{2}{*}{$_{\Lambda}^{10}$Be}
     & \multirow{2}{*}{8.508}  & \multirow{4}{*}{0.235} & \multirow{2}{*}{0.278}  & \multirow{2}{*}{9.006}   & \multirow{4}{*}{0.220 }& \multirow{4}{*}{0.217 } & \multirow{2}{*}{$0.22\pm0.25$} & $9.11\pm0.22$~\cite{cantwell1974binding}\\

&                                  
    &   &  &  &  & && ~\cite{cantwell1974binding,davis200550} & $8.60\pm0.07$~\cite{gogami2016high}\\

& \multirow{2}{*}{$_{\Lambda}^{10}$B}  
   & \multirow{2}{*}{8.273}  & & \multirow{2}{*}{0.279}   & \multirow{2}{*}{8.786}   & & & \multirow{2}{*}{$-0.04\pm0.12$}& $8.89\pm0.12$~\cite{davis200550}\\  
&  &   &  &  &  & &&~\cite{gogami2016high} & $8.64\pm0.1$~\cite{gogami2016high}\\

\midrule
\multirow{2}{*}{$A=12$} 
& $_{\Lambda}^{12}$B  
  & 10.857  & \multirow{2}{*}{0.233} & 0.109  & 11.088   & \multirow{2}{*}{0.225}& \multirow{2}{*}{0.229} & \multirow{2}{*}{$0.23\pm0.19$} & $11.529\pm0.025$~\cite{tang2014experiments}\\
& $_{\Lambda}^{12}$C       & 10.024   & 
       & 0.113  & 10.863   &  &                    && $11.30\pm0.19$~\cite{dluzewski1988binding,davis200550,gogami2016high} \\
       
\midrule
\multirow{2}{*}{$A=16$} 
& $_{\Lambda}^{16}$N  
 & 13.707  & \multirow{2}{*}{0.148}& \multirow{2}{*}{0}  & 13.707   & \multirow{2}{*}{0.148} & \multirow{2}{*}{-} & \multirow{2}{*}{$0.36\pm0.43$} & $13.76\pm0.16$~\cite{cusanno2009high}\\
& $_{\Lambda}^{16}$O  & 13.559   & 
       &   & 13.559   &       &               & & $13.4\pm0.4$~\cite{agnello2011hypernuclear}\\

\midrule
\multirow{2}{*}{$A=32$} 
& $_{\Lambda}^{32}$P & 18.832  & \multirow{2}{*}{0.125} & \multirow{2}{*}{0}  & 18.832   & \multirow{2}{*}{0.125} & \multirow{2}{*}{-} & \multirow{2}{*}{-}& \multirow{2}{*}{-}  \\
& $_{\Lambda}^{32}$S       & 18.707   & 
       &   & 18.707   &         &             & &\\       
       
\midrule
\multirow{2}{*}{$A=40$}
& $_{\Lambda}^{40}$K      & 19.976   & \multirow{2}{*}{0.119} 
         & 0.019   & 20.079                 & \multirow{2}{*}{0.067} & \multirow{2}{*}{0.089} & \multirow{2}{*}{-} & -\\
& $_{\Lambda}^{40}$Ca      & 19.857  && 0.025  & 20.012   & &&& $18.7\pm1.1$~\cite{pile1991study} \\

\bottomrule
\end{tabular}
\end{table*}

\subsection{CSB effects in Hypernuclei}
We performed calculations using the DSHF + BCS approach, with the CSB strength $v_{0}$ which is optimized to fit the experimental data of the difference in the single-$\Lambda$ binding energies of six pairs of mirror hypernuclei ($^{7}_{\Lambda}$He, $^{7}_{\Lambda}$Be), 
($^{8}_{\Lambda}$Li, $^{8}_{\Lambda}$Be), ($^{9}_{\Lambda}$Li, $^{9}_{\Lambda}$B), ($^{10}_{\Lambda}$Be, $^{10}_{\Lambda}$B), 
($^{12}_{\Lambda}$B, $^{12}_{\Lambda}$C), ($^{16}_{\Lambda}$N, $^{16}_{\Lambda}$O).
The extracted CSB strengths are  $v_{0} = 27.78 \text{ MeV fm}^{-3}$
 and $v_{0} = 27.39 \text{ MeV fm}^{-3}$, for the $NN$ interactions SLy4 and SLy5, respectively.\par
 
 Fig.~\ref{plt.Pjsp} shows the single-$\Lambda$ binding energy differences between mirror hypernuclei (add all the calculated hypernuclei) with or without $\Lambda N$ CSB interaction and hypernuclei deformations, in the case of $NN$ interaction SLy4 (upper panle) and SLy5 (lower panel), respectively. For comparison, the experimental data and the RMF resluts are given. We can see that the calculated $\Delta B_{\Lambda}$ with CSB and experimental values for the hypernuclei pairs $A = 10$ and $A = 12$ overlap highly. Compared with all the available experimental values, namely the hypernuclei pairs $A = 7-10, 12, 16$, the calculated results with the CSB interaction show a significant improvement in $\Delta B_\Lambda$ than those without the CSB term. The RMF model (blue open squares) with CSB interaction yields binding energy differences for four mirror hypernuclei pairs. Comparing the two methods, there are disparities in their calculated values, and the DSHF results exhibit distinct characteristics relative to those from the RMF model.
 
 Additionally, our predicted results for $A = 32$ and $ A = 40$ show a significant increase compared to those without CSB. The result for the hypernuclei pair with $A = 12$ deviates somewhat from the experimental value but still lies within the error range. They also predict results for the hypernuclei pair with $A = 40$, which show an increase compared to those without CSB.\par
 
Table~\ref{sly4} and Table~\ref{sly5} presents single-$\Lambda$ binding energies $B_\Lambda$, binding energy differences $\Delta B_{\Lambda(\text{SHF})}$ (from DSHF + BCS with/without $ \Lambda N$ CSB interaction) and $\Delta B_{\Lambda(\text{RMF})}$ ~\cite{sun2025charge}, quadrupole deformation parameter $\beta_2$, and available experimental $B_\Lambda$ and $\Delta B_{\Lambda(\text{Expt.})}$ data for eight pairs of mirror hypernuclei. We also incorporated the results of previous theoretical studies for the ease of comparison. The no-core shell model employ the NLO13 and NLO19 potentials to study CSB in the $A=8$ isodoublet~\cite{le2023ab}. Gal~\cite{gal2015charge} computed the CSB effect by employing a shell-model approach in combination with an effective $\Lambda\Sigma$ coupling model. The $A = 8$ calculation by Hiyama et al.~\cite{hiyama2009} was done within a $(\Lambda + \alpha + ^3\text{He}/t)$ three-body cluster cluster model. \par
For $A=7$ hypernuclei pair, $\Delta B_\Lambda$ is only 0.044-0.061 MeV without CSB, increasing to 0.431-0.436 MeV when CSB is included, which agrees well with the experimental value of $0.39 \pm 0.13$ MeV. Using the $NN$ interaction SLy4 and SLy5 with CSB, we obtain the CSB effects of 0.530 MeV and 0.539 MeV for $^{9}_{\Lambda}$Li-$^{9}_{\Lambda}$B pair,  respectively. This is well in line with the empirical CSB of $0.65\pm0.21$ MeV~\cite{jurivc1973new}, which is extracted from the separation energies determined by emulsion experiments. Our prediction for
 ($^{10}_{\Lambda}$Be, $^{10}_{\Lambda}$B) using $NN$ interaction SLy4(SLy5) with CSB interaction, $\Delta B_\Lambda$ = 0.220 MeV, is identical to the CSB estimated by the RMF model~\cite{sun2025charge}. Meanwhile Gal~\cite{gal2015charge} gives 0.136 MeV, showing a consistent trend with our model regarding the influence of core deformation.\par 
Moreover, both $A=16$ and $A=32$ hypernuclei are spherical. For $A=32$, $\Delta B_{\Lambda(\text{SHF})}$ increases from 0.009-0.010 MeV to 0.125 MeV due to CSB, demonstrating a strong CSB enhancement effect. For $A=16$, SHF calculations yield 0.144-0.148 MeV, which falls within the experimental range of $0.36 \pm 0.43$ MeV, demonstrating that the model can be used to study large-mass nuclei.\par
CSB results for the $A=8$ mirror hypernuclei are listed at the upper part of Table~\ref{sly4} and Table~\ref{sly5}.When using the $NN$ interaction SLy4 and SLy5 without CSB, a negligibly small CSB is predicted for $^{8}_{\Lambda}$Li-$^{8}_{\Lambda}$Be, namely
 $\Delta B_\Lambda$ = 0.063 MeV and 0.059 MeV, respectively. This is, however, well in line with the empirical CSB of $-0.040\pm0.060$ MeV~\cite{jurivc1973new} extracted from the separation energies determined by emulsion experiments. A similarly small $\Delta B_\Lambda$ was also predicted by Gal in~\cite{gal2015charge}, $\Delta B_\Lambda$ = -0.049 MeV. With the CSB interaction included, both the $NN$ interaction SLy4 and SLy5 yield sizable CSB results,
 $\Delta B_\Lambda$ = 0.256 MeV and 0.254 MeV.  Interestingly, the predictions of NLO13 and NLO19 with CSB interaction are -0.178 MeV and -0.143 MeV, which is  comparable to the value of $\Delta B_\Lambda$ = -0.160 MeV obtained by Hiyama et al.~\cite{hiyama2009}. For $A=12$ hypernuclei, the calculated $\Delta B_{\Lambda(\text{SHF})} = 0.225$ MeV deviates by only 0.005 MeV from the experimental value of $0.23 \pm 0.19$ MeV, indicating high model precision. For heavy $A=40$ nuclei, the deviation between SHF and RMF calculations is approximately 0.03 MeV, suggesting weak deformation effects and good model consistency in heavy nuclei. This CSB-driven enhancement of $\Delta B_{\Lambda(\text{SHF})}$ is observed in most hypernuclei mirror pairs, effectively improving agreement between theoretical calculations and experimental results.\par

\subsection{Deformation effects on CSB}
Fig.~\ref{deformed} presents the calculated single-$\Lambda$ binding energies $B_{\Lambda}$ and their differences $\Delta B_{\Lambda}$, encompassing calculations from the DSHF + BCS approach (with CSB interaction, both with and without nuclear deformation) and the RMF model (with CSB interaction but without nuclear deformation), along with experimental data, for various hypernuclei systems, and where the upper and lower panels correspond to results with $NN$ interaction SLy4 and SLy5, respectively.\par 

For the SHF results, the calculated values of the $A = 7$ hypernuclei pair, considering deformation leads to improved results. For $A = 9\sim12$, the results with deformation are significantly smaller than those without. For $A = 10$ and $A = 12$, when nuclear deformation is taken into account, highly overlap with the experimental values. Additionally, the predicted results for $A = 32$ and $A = 40$ exhibit notable enhancement compared to the cases without the CSB interaction.For $A = 7$ and $A = 10$ hypernuclei pairs, RMF results show good consistency with SHF results (with deformation and CSB). For $A = 12$, RMF results deviate slightly from SHF ones but still fall within reasonable ranges, indicating some differences in model predictions. \par

Table~\ref{tab:deformation} presents single-$\Lambda$ binding energies $B_\Lambda$, binding energy differences $\Delta B_{\Lambda(\text{SHF})}$ (calculated with $NN$ interaction SLy4, with and without nuclear deformation and with CSB), quadrupole deformation parameter $\beta_2$, binding energy differences $\Delta B_{\Lambda(\text{RMF})}$ from the RMF model, and available experimental $B_\Lambda$ and $\Delta B_{\Lambda(\text{Expt.})}$ data for various hypernuclei systems. \par
For the $A=7$ pair with minimal deformation ($\beta_2 = 0.009$ and $0.001$), $\Delta B_{\Lambda(\text{SHF})}$ slightly increases from $0.423$ MeV to $0.436$ MeV, enhancing the CSB effect but slightly enlarging the deviation from the experimental value ($0.39 \pm 0.13$ MeV), while the RMF result (0.368 MeV) aligns better with experiments in Ref.~\cite{sun2025charge}. The $A=16$ pair, being spherical or almost spherical, has $\Delta B_{\Lambda(\text{SHF})}$ remain unchanged at $0.148$ MeV whether deformation is included or not, confirming CSB insensitivity to deformation here. The $A=32$ pair, also nearly spherical, sees its $\Delta B_{\Lambda(\text{SHF})}$ stay at $0.125$ MeV with or without deformation.\par 
In systems with notable deformation, for the $A=8$ pair with a large quadrupole deformation ($\beta_2 = 0.297$ and $0.324$), $\Delta B_{\Lambda(\text{SHF})}$ is notably modulated from $0.283$ MeV to $0.256$ MeV. For the $A=9$ pair, the mirror pair shows different types of deformation. $^9_\Lambda\mathrm{Li}$ is an oblate one ($\beta_2 = -0.164$) and $^9_\Lambda\mathrm{B}$ is a prolate one ($\beta_2 = 0.284$). Correspondingly, $\Delta B_{\Lambda(\text{SHF})}$ changes distinctly from $0.604$ MeV to $0.530$ MeV, bringing the calculated results slightly closer to experiments. The $A=10$ pair with comparable deformation ($\beta_2 = 0.278$ and $0.279$) leads to a moderate $\Delta B_{\Lambda(\text{SHF})}$ reduction from $0.235$ MeV to $0.220$ MeV, improving agreement with experiments, and the RMF calculation gives $\Delta B_{\Lambda(\text{RMF})} = 0.217$ MeV in Ref.~\cite{sun2025charge}. The $A=12$ pair with small $\beta_2$ ($0.109$ and $0.113$) results in minimal $\Delta B_{\Lambda(\text{SHF})}$ variation from $0.233$ MeV to $0.225$ MeV, consistent with the experimental value ($0.23 \pm 0.19$ MeV). For the heavy $A=40$ pair, $\Delta B_{\Lambda(\text{RMF})} = 0.089$ MeV, while DSHF gives $0.067$ MeV (with deformation) and $0.119$ MeV (without deformation), which shows that $\Delta B_{\Lambda}$ shows model dependence. All results have small $\Delta B_{\Lambda}$ values, consistent with the weak CSB effect in heavy nuclei. Overall, deformation notably alters $B_{\Lambda}$ and $\Delta B_{\Lambda}$, especially in light and mid-mass hypernuclei with pronounced quadrupole deformation.\par
By comparing the results with and without deformation, we find that deformation leads to noticeable modifications in $B_{\Lambda}$ and $\Delta B_{\Lambda}$ for several hypernuclei, particularly in light and mid-mass systems where quadrupole deformation is more pronounced.\par

\section{Summary}
\label{sec4}
We studied the CSB effect found in the single-$\Lambda$ binding energy difference of mirror hypernuclei in the mass region of $A=7\sim 40$ by using the DSHF model with Skyrme EDFs SLy4 and SLy5 adding the CSB terms in the Hamiltonian. The strength of the CSB term is optimized to fit all six sets of experimental CSB data in the mirror hypernuclei. Our model describes reasonably well experimental single-$\Lambda$ binding energy differences of mirror hypernuclei with masses from $A=7\sim 16$. To check the dependence on the EDF model, we employed two different EDFs (SLy4 and SLy5) in the calculations. Essentially identical results are find in the two panels of Fig.~\ref{plt.Pjsp}. These results demonstrate that the CSB term plays a crucial role in describing the ISB effect in hypernuclei systems.\par
In the study, the inclusion of the CSB effect significantly increases the single-$\Lambda$ binding energy difference $\Delta B_{\Lambda}$ for most mirror hypernuclei pairs, with the calculated results for $A=7$, $A=10$, and $A=12$ showing excellent agreement with experimental values, and those for $A=9$ and $A=16$ falling well within the experimental error range, thus effectively improving the consistency between theory and experiment. Meanwhile, the introduction of the CSB effect exhibits consistent trends across the DSHF+BCS and RMF models, particularly evident in $A=7$, $A=10$, and $A=12$ systems, verifying its crucial role in describing charge symmetry breaking in hypernuclei.\par
The effect of considering deformation on hypernuclei $B_{\Lambda}$ and $\Delta B_{\Lambda}$ is related to the hypernuclei mass and shape. In light-mass hypernuclei with more pronounced quadrupole deformation ($A=8$, $A=9$), deformation significantly modulates $\Delta B_{\Lambda}$ and mostly brings calculations closer to experimental values, while in nearly spherical hypernuclei ($A=16$, $A=32$), the effect of deformation is minimal.
In the heavy nucleus $A=40$, the overall $\Delta B_{\Lambda}$ value is small, and the significant effect of deformation is mainly concentrated in light/medium-mass hypernuclei with more prominent quadrupole deformation.\par 
We also predict the single-$\Lambda$ binding energy differences of medium-heavy nuclei with $A=32$ and 40 for further confirmation of the effect.This prediction can be verified by future heavy-ion collision experiments (RHIC-STAR, JLab CLAS12), which will provide crucial evidence for investigating the mass dependence of the CSB effect in heavy hypernuclei.

\section*{Acknowledgments}
This work was supported by the National Natural Science Foundation of China under Grants No. 12175071 and No.
 12575124, the Fundamental Research Funds for the Central Universities.


\newcommand{\epja}{EPJA}
\newcommand{\npa}{Nucl. Phys. A}
\newcommand{\nphysa}{Nucl. Phys. A}
\newcommand{\ppnp}{Prog. Part. Nucl. Phys.}
\newcommand{\ptp}{Prog. Theor. Phys.}
\newcommand{\ptep}{Prog. Theor. Exp. Phys.}
\bibliography{csb}
\end{document}